\documentclass[conference,letterpaper,twoside,twocolumn,final]{IEEEtran}
\usepackage[cmex10]{amsmath}
\usepackage{amssymb,latexsym}
\usepackage{color}
\usepackage{graphicx}
\usepackage{float}
\usepackage{psfrag}
\usepackage{cancel}
\usepackage[{Procedure}]{algorithm}
\usepackage{algpseudocode}
\usepackage{enumerate}

\usepackage{tikz}
\usetikzlibrary{patterns}

\newlength{\hatchspread}
\newlength{\hatchthickness}
\newlength{\hatchshift}
\newcommand{\hatchcolor}{}
\tikzset{hatchspread/.code={\setlength{\hatchspread}{#1}},
         hatchthickness/.code={\setlength{\hatchthickness}{#1}},
         hatchshift/.code={\setlength{\hatchshift}{#1}},
         hatchcolor/.code={\renewcommand{\hatchcolor}{#1}}}
\tikzset{hatchspread=3pt,
         hatchthickness=0.4pt,
         hatchshift=0pt,
         hatchcolor=black}
\pgfdeclarepatternformonly[\hatchspread,\hatchthickness,\hatchshift,\hatchcolor]
   {custom north west lines}
   {\pgfqpoint{\dimexpr-2\hatchthickness}{\dimexpr-2\hatchthickness}}
   {\pgfqpoint{\dimexpr\hatchspread+2\hatchthickness}{\dimexpr\hatchspread+2\hatchthickness}}
   {\pgfqpoint{\dimexpr\hatchspread}{\dimexpr\hatchspread}}
   {
    \pgfsetlinewidth{\hatchthickness}
    \pgfpathmoveto{\pgfqpoint{0pt}{\dimexpr\hatchspread+\hatchshift}}
    \pgfpathlineto{\pgfqpoint{\dimexpr\hatchspread+0.15pt+\hatchshift}{-0.15pt}}
    \ifdim \hatchshift > 0pt
      \pgfpathmoveto{\pgfqpoint{0pt}{\hatchshift}}
      \pgfpathlineto{\pgfqpoint{\dimexpr0.15pt+\hatchshift}{-0.15pt}}
    \fi
    \pgfsetstrokecolor{\hatchcolor}
    \pgfusepath{stroke}
   }

\newlength{\starsize}
\newlength{\starspread}
\tikzset{starsize/.code={\setlength{\starsize}{#1}},
         starspread/.code={\setlength{\starspread}{#1}}}
\tikzset{starsize=1mm,
         starspread=3mm}
\pgfdeclarepatternformonly[\starspread,\starsize]
  {custom fivepointed stars}
  {\pgfpointorigin}
  {\pgfqpoint{\starspread}{\starspread}}
  {\pgfqpoint{\starspread}{\starspread}}
  {
   \pgftransformshift{\pgfqpoint{\starsize}{\starsize}}
   \pgfpathmoveto{\pgfqpointpolar{18}{\starsize}}
   \pgfpathlineto{\pgfqpointpolar{162}{\starsize}}
   \pgfpathlineto{\pgfqpointpolar{306}{\starsize}}
   \pgfpathlineto{\pgfqpointpolar{90}{\starsize}}
   \pgfpathlineto{\pgfqpointpolar{234}{\starsize}}
   \pgfpathclose%
   \pgfusepath{fill}
  }

\graphicspath{ {./figures/} }

\newcommand{\ScPSF}{1} 
\newcommand{\gridw}{0.4}
\newcommand{\gridh}{0.075}

\newcommand{\herm}[1]{{#1}^{\mathsf{H}}}
\newcommand{\figref}[1]{Fig.~\ref{#1}}

\usepackage{mathtools}
\DeclarePairedDelimiter{\ceil}{\lceil}{\rceil}
\DeclarePairedDelimiter{\floor}{\lfloor}{\rfloor}

\usepackage{silence}
\usepackage[acronym]{glossaries}
\newacronym{OFDM}{OFDM}{orthogonal frequency-division multiplexing}
\newacronym{MAC}{MAC}{medium access control}
\newacronym{PHY}{PHY}{physical}
\newacronym{CP}{CP}{cyclic prefix}
\newacronym{CRC}{CRC}{cyclic redundancy check}
\newacronym{LS}{LS}{least-squares}
\newacronym{FER}{FER}{frame error rate}
\newacronym{MCS}{MCS}{modulation and coding scheme}
\newacronym{LLR}{LLR}{log-likelihood ratio}

\begin{document}
\title{On Channel Estimation for 802.11p in Highly Time-Varying Vehicular Channels\textsuperscript{\small *}\vspace{-3ex}}\author{\IEEEauthorblockN{Keerthi Kumar Nagalapur, Fredrik Br\"annstr\"om, and Erik G. Str\"om}
\IEEEauthorblockA{Department of Signals and Systems, Chalmers University of Technology, Gothenburg, Sweden \\
\emph{\{keerthi, fredrik.brannstrom, erik.strom\}@chalmers.se}}}
\maketitle

\begin{abstract}
Vehicular wireless channels are highly time-varying and the pilot pattern in the 802.11p orthogonal frequency-division multiplexing frame has been shown to be ill suited for long data packets. The high frame error rate in off-the-shelf chipsets with noniterative receiver configurations is mostly due to the use of outdated channel estimates for equalization. This paper deals with improving the channel estimation in 802.11p systems using a cross layered approach, where known data bits are inserted in the higher layers and a modified receiver makes use of these bits as training data for improved channel estimation. We also describe a noniterative receiver configuration for utilizing the additional training bits and show through simulations that frame error rates close to the case with perfect channel knowledge can be achieved.
\end{abstract}

\renewcommand*{\thefootnote}{\fnsymbol{footnote}}
\footnotetext[1]{This is a revised version of the article found in IEEEXplore. The revision is related to one of the results in Fig. 11 and its interpretation.}

\section{Introduction}
\let\thefootnote\relax\footnote{The research was partially funded by Chalmers Antenna Systems Excellence Center in the project `Antenna Systems for V2X Communication'.}
IEEE 802.11p, whose \gls{PHY} layer is identical with the widely used 802.11a, has been chosen as the standard for direct vehicle-to-vehicle communication. In 802.11p a default channel spacing of $10$ MHz is proposed as opposed to the commonly used $20$ MHz spacing in 802.11a to provide better protection against longer delay spreads. The length of the \gls{CP} and subcarrier spacing satisfy the design guidelines of an \gls{OFDM} system for the reported values of delay and Doppler spreads in vehicular channel measurements~\cite{Bernado2013}. However, the training data in 802.11p originally intended for 802.11a is designed for relatively static devices and indoor use. In vehicular communications high vehicular velocities, dynamic environment, and long packet sizes of around $400$ bytes mean that the channel frequency response can change significantly within one \GLS{OFDM} frame~\cite{Bernado2013}. Robust channel estimates for the whole frame are required to ensure the low \glspl{FER} that traffic safety applications desire.

To address the high \gls{FER} due to nonrobust channel estimation, solutions using advanced receivers and channel tracking have been proposed. In~\cite{zemen2012} a receiver with iterative channel estimation has been proposed that achieves FER performance of a receiver with perfect channel knowledge. Also, insertion of a postamble in the \gls{PHY} layer to reduce the complexity of the iterative channel estimation is proposed. The advantage of periodically inserted training \gls{OFDM} symbols referred to as midambles has been shown in~\cite{Woong2009} and gain in bit error rate performance has been reported.

In this paper, we develop a detailed procedure to add additional training symbols to an 802.11p \gls{OFDM} frame. We accomplish this, in contrast to previous works, e.g.,~\cite{Woong2009} without modifying the transmitter \gls{PHY} or \gls{MAC} layers. A modified receiver can utilize these known additional symbols to improve the channel estimation. A standard receiver sees the additional training symbols as part of the data and passes it to the higher layers (layers above the \gls{PHY} and \gls{MAC} layers), where the additional bits corresponding to the inserted symbols can be removed. A standard receiver requires a simple software update in the higher layers to discard the additional data, making our scheme \emph{backward compatible with the added requirement of a software update}. We also describe an implementation of the receiver which uses the inserted training symbols for robust channel estimation. A modified 802.11p frame (MF) with inserted training bits can be indicated by using the reserved bit in the SIGNAL field of the 802.11p frame~\cite[Sec. 18.3]{IEEE802112012}.


To the best of our knowledge the \gls{MAC} and \gls{PHY} layer functionalities of an 802.11p transceiver are implemented in the chip and the layers above them are software defined, which can be modified with a software update. Also, since 802.11p uses ``outside the context of a basic service set communication'', encryption and authentication are not performed by the \gls{MAC} layer~\cite{IEEE80211p}. As a consequence the data from the layer above the \gls{MAC} layer enters the \GLS{PHY} layer unchanged with the insertion of headers and \gls{CRC} checksum of fixed lengths.

The main contributions of the paper are: (i) We propose a cross layered approach to insert more training data before the \gls{PHY} and \gls{MAC} layers into an 802.11p frame; (ii) A receiver configuration to utilize the inserted training data to obtain robust channel estimates is described and its performance is evaluated using simulations.

\section{System Model and {802.11p} Frame Structure}

The \gls{PHY} layer of 802.11p uses \GLS{OFDM} with $N=64$ subcarriers and a \gls{CP} of length $N_{\rm{CP}}=16$. Among the $64$ subcarriers, $48$ are allocated for data, $4$ are allocated for pilots and $12$ are null subcarriers. A channel spacing of $10$ MHz results in \GLS{OFDM} symbol duration of $T_{\rm{SYM}} = 8 \, \rm{\mu}s$ which includes a \gls{CP} of $1.6 \, \rm{\mu}s$. The 802.11p standard supports eight different \glspl{MCS}~\cite[Table 18.4]{IEEE802112012}.

\begin{figure*}[!t]
\renewcommand{\ScPSF}{0.8}
\psfrag{Sc}  [c][c][\ScPSF]{Scrambler}%
\psfrag{En}  [c][c][\ScPSF]{Encoder}%
\psfrag{In}  [c][c][\ScPSF]{Interleaver}%
\psfrag{Ma}  [c][c][\ScPSF]{Mapper}%
\psfrag{SP}  [c][c][\ScPSF]{S/P}%
\psfrag{IFFT}[c][c][\ScPSF]{IDFT}%
\psfrag{CP}  [c][c][\ScPSF]{CP}%
\psfrag{PS}  [c][c][\ScPSF]{P/S}%
\psfrag{C}   [c][c][\ScPSF]{}
\psfrag{P}   [c][c][\ScPSF]{$P[m,k]$}%
\psfrag{S}   [c][c][\ScPSF]{$S[m,k]$}%
\psfrag{s}   [c][c][\ScPSF]{$s[m,n]$}%
\psfrag{sn}  [c][c][\ScPSF]{}
\centering
\includegraphics[width=0.9\linewidth]{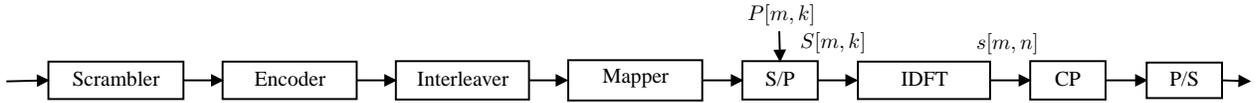}
\caption{Block diagram of an 802.11p transmitter.}
\label{fig:trans_block}
\end{figure*}

The encoding process of an 802.11 \GLS{OFDM} symbol is illustrated in \figref{fig:trans_block}. Data bits are scrambled and encoded using a rate $1/2, \, (171, 133)_8$ convolutional code. Higher coding rates are achieved using puncturing. The encoded bits are divided into groups of $N_{\rm{CBPS}}$ bits, where $N_{\rm{CBPS}}$ is the number of coded bits per \GLS{OFDM} symbol. The bits in each group are interleaved and mapped to the constellation $\mathcal{C}$ resulting in complex valued symbols. The complex valued symbols and the pilots are mapped to the data subcarriers and the pilot subcarriers of the \GLS{OFDM} symbol, respectively. Following which, an $N$-point inverse discrete Fourier transform (IDFT) is performed to obtain the time domain signal and a \gls{CP} of $N_{\rm{CP}}$ samples is added to form the \GLS{OFDM} symbol. For specific details of the various blocks in the transmitter see~\cite[Ch. 18]{IEEE802112012}.

Assigning the data symbol positions to the set $\mathcal{D}$ and the pilot positions to the set $\mathcal{P}$, the frequency domain symbol mapped to the $k$th subcarrier in the $m$th \GLS{OFDM} symbol is given by
\begin{equation}
S[m,k] = D[m,k] + P [m,k],
\end{equation}
where $D[m,k]$ and $P[m,k]$ are the data and the pilot symbols respectively, such that $P[m,k]=0, \: \forall[m,k] \in \mathcal{D}$ and $D[m,k]=0, \: \forall[m,k] \in \mathcal{P}$. The time domain samples obtained after the $N$-point IDFT are denoted by $s[m,n]$.


\subsection{{802.11p} Frame} 
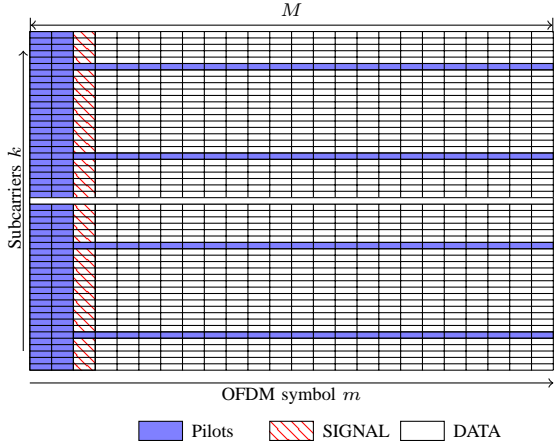
\begin{figure}[!t]
\centering
\begin{tikzpicture}
\renewcommand{\gridw}{0.29}
\renewcommand{\gridh}{0.085}
\filldraw[fill=blue!50!white]                                             (0,0)        rectangle (2*\gridw,53*\gridh);
\filldraw[pattern=custom north west lines,hatchspread=4pt,hatchcolor=red] (2*\gridw,0) rectangle (3*\gridw,53*\gridh);
\filldraw[fill=blue!50!white] (2*\gridw,5*\gridh)  rectangle (24*\gridw,6*\gridh);
\filldraw[fill=blue!50!white] (2*\gridw,19*\gridh) rectangle (24*\gridw,20*\gridh);
\filldraw[fill=blue!50!white] (2*\gridw,33*\gridh) rectangle (24*\gridw,34*\gridh);
\filldraw[fill=blue!50!white] (2*\gridw,47*\gridh) rectangle (24*\gridw,48*\gridh);
\draw [draw=black, xscale=\gridw, yscale=\gridh] (0,0)  grid (24,53);
\filldraw[fill=white, opacity=1] (0*\gridw,26*\gridh) rectangle (24*\gridw,27*\gridh);
\draw[fill=blue!50!white]                                            (5*\gridw,-11*\gridh) rectangle (7*\gridw,-8*\gridh) (7*\gridw,-9.5*\gridh)   node[align=center,right]{\scriptsize Pilots} ;
\draw[pattern=custom north west lines,hatchspread=4pt,hatchcolor=red](11*\gridw,-11*\gridh) rectangle(13*\gridw,-8*\gridh)(13*\gridw,-9.5*\gridh)   node[align=center,right]{\scriptsize SIGNAL};
\draw[fill=white,          opacity=1]                               (17*\gridw,-11*\gridh) rectangle (19*\gridw,-8*\gridh)(19*\gridw,-9.5*\gridh) node[align=center,right]{\scriptsize DATA};
\draw [->] (-0.3*\gridw,3*\gridh) -- (-0.3*\gridw,50*\gridh);
\draw [|<->|] (0,54*\gridh) -- (24*\gridw,54*\gridh) node[midway,above]{\scriptsize $M$} ;
\node[align=center,rotate=90] at (-0.7*\gridw,27*\gridh){\scriptsize Subcarriers $k$} ;
\draw [->] (0*\gridw,-2*\gridh) -- (24*\gridw,-2*\gridh);
\node[align=center,rotate=0] at (12*\gridw,-4*\gridh){\scriptsize OFDM symbol $m$} ;
\end{tikzpicture}
\caption{A standard 802.11p frame in subcarrier-time grid showing the position of the pilots and the data symbols.}
\label{fig:80211frame}
\end{figure}

\figref{fig:80211frame} shows a standard 802.11p frame (SF) in a subcarrier-time grid. A frame begins with two identical \GLS{OFDM} symbols referred to as long training (LT) symbols. The SIGNAL symbol carries the information regarding the length of the packet and the \gls{MCS} used. The SIGNAL symbol is always encoded using the rate $1/2, \, (171, 133)_8$ convolutional code without puncturing together with BPSK. The SIGNAL field is followed by the \GLS{OFDM} symbols carrying the data. Each of the data OFDM symbols includes four pilots referred to as \emph{comb pilots}. A sequence of 10 identical short symbols spanning over a duration of $2 \cdot T_{\rm{SYM}}$, referred to as short training, is prefixed to the frame (not shown in the figure). This sequence is used in the receiver for signal detection and synchronization. The number of OFDM symbols in the SF beginning with the two LT symbols is denoted by $M$.

\section{Modified Frame}
\begin{figure}[!t]
\renewcommand{\ScPSF}{0.6}
\psfrag{b16} [c][c][\ScPSF]{$N_{\mathrm{SERV}}$}
\psfrag{N}   [c][c][\ScPSF]{$N_{\mathrm{MACH}}$}
\psfrag{NFB} [c][c][\ScPSF]{$N_{\mathrm{FB}}$}
\psfrag{Nc}  [c][c][\ScPSF]{$N_{\mathrm{CRC}}$}
\psfrag{Nt}  [c][c][\ScPSF]{$N_{\mathrm{MEM}}$}
\psfrag{MH}  [c][c][\ScPSF]{MAC Header}
\psfrag{FB}  [c][c][\ScPSF]{FB}
\psfrag{CRC} [c][c][\ScPSF]{CRC}
\psfrag{SER} [c][c][\ScPSF]{SERVICE}
\psfrag{PSDU}[c][c][\ScPSF]{PSDU}
\psfrag{TAIL}[c][c][\ScPSF]{TAIL}
\psfrag{PAD} [c][c][\ScPSF]{PAD}
\psfrag{LT1} [c][c][\ScPSF]{LT1}
\psfrag{LT2} [c][c][\ScPSF]{LT2}
\psfrag{SIGNAL} [c][c][\ScPSF]{SIGNAL}
\psfrag{DATA}[c][c][\ScPSF]{DATA}
\psfrag{LLC} [r][c][\ScPSF]{LLC:}
\psfrag{MAC} [r][c][\ScPSF]{MAC:}
\psfrag{PLCP}[r][c][\ScPSF]{PLCP:}
\psfrag{PHY} [r][c][\ScPSF]{PHY:}
\centering
\includegraphics[width=\linewidth]{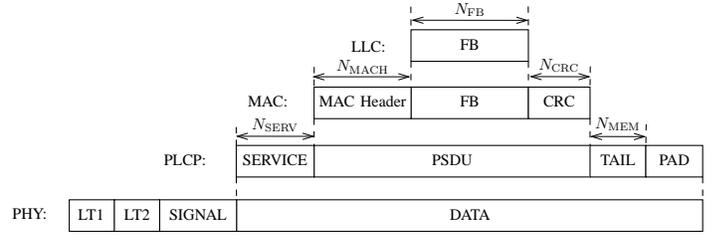}
\caption{Diagram showing the flow of data bits from LLC sublayer to the \GLS{PHY} layer.}
\label{fig:dataflow}
\end{figure}

In this section the proposed MF and its encoding process is described. The flow of the data from the logical link control (LLC) sublayer to the \GLS{PHY} layer is shown in \figref{fig:dataflow}. The LLC sublayer outputs a data unit referred to as the Frame Body (FB) of length $N_{\rm{FB}}$ bits. The FB is passed to the \gls{MAC} layer, which adds a \gls{MAC} header of length $N_{\rm{MACH}} = 288$ bits and a \gls{CRC} checksum of length $N_{\rm{CRC}} = 32$ bits computed over the \gls{MAC} header and the FB to form the \gls{PHY} layer convergence protocol (PLCP) service data unit (PSDU). In the PLCP sublayer, a SERVICE field of $N_{\rm{SERV}}=16$ bits, a TAIL of $N_{\rm{MEM}}=6$ zero bits, and padded zero bits are added to form the DATA unit of the \GLS{OFDM} frame. Since the \gls{MAC} layer in 802.11p does not perform any encryption, the location of the FB bits in the DATA unit can be determined. \emph{The insertion of the additional training bits in the FB to form the MF is performed immediately after the LLC sublayer (before passing it down to the MAC layer), under the assumption that the LLC sublayer is software-defined, this is easily done with a software update.}

The training bits inserted in the LLC sublayer are scrambled in the \GLS{PHY} layer. Since the scrambler is initialized with a pseudo-random seed, the training bits inserted in the LLC sublayer are modified in a manner unknown to the LLC sublayer. However, the pseudo-random sequence used to initialize the scrambler is estimated in the receiver. The scrambler seed in combination with the training data bits inserted in the transmitter can be used by the receiver to recreate the output of the scrambler.

The rate $1/2, \, (171, 133)_8$ convolutional encoder has a memory of $N_{\rm{MEM}}$ bits. As a consequence, a bit that is the output of the convolutional encoder is a function of the input bit and $N_{\rm{MEM}}$ previous input bits. To insert known training bits into the frame, the convolutional encoder is terminated in a known state before the training bits are fed to the convolutional encoder. The resulting output bits are therefore determined by the termination state and the input bits. Hence, inserting a known training sequence of any length into the frame requires $N_{\rm{MEM}}$ additional bits. Also, the encoded bits are interleaved over one OFDM symbol modifying the bits positions. To minimize the overhead of termination and to insert training data which facilitates robust channel estimation for the frame, we insert training bits of length $N_{\mathrm{DBPS}}$ (number of data bits per \GLS{OFDM} symbol) periodically in the frame. Inserting a complete \GLS{OFDM} training symbol requires the convolutional encoder to be terminated in a known state only once per training symbol resulting in smaller overhead. In addition, a complete \GLS{OFDM} symbol as training, similar to an LT symbol, has the advantage of measuring the frequency response across the whole signal bandwidth.

\begin{figure}
\centering
\begin{tikzpicture}
\renewcommand{\gridw}{0.29}
\renewcommand{\gridh}{0.085}
\filldraw[fill=blue!50!white]                                             (0,0)        rectangle (2*\gridw,53*\gridh);
\filldraw[pattern=custom north west lines,hatchspread=4pt,hatchcolor=red] (2*\gridw,0) rectangle (3*\gridw,53*\gridh);
\filldraw[fill=blue!50!white] (10*\gridw,0) rectangle (11*\gridw,53*\gridh);
\filldraw[fill=blue!50!white] (15*\gridw,0) rectangle (16*\gridw,53*\gridh);
\filldraw[fill=blue!50!white] (20*\gridw,0) rectangle (21*\gridw,53*\gridh);
\filldraw[fill=blue!50!white] (23*\gridw,0) rectangle (24*\gridw,53*\gridh);
\filldraw[fill=blue!50!white] (26*\gridw,0) rectangle (27*\gridw,53*\gridh);
\filldraw[fill=blue!50!white] (2*\gridw,5*\gridh)  rectangle (21*\gridw,6*\gridh);
\filldraw[fill=blue!50!white] (2*\gridw,19*\gridh) rectangle (21*\gridw,20*\gridh);
\filldraw[fill=blue!50!white] (2*\gridw,33*\gridh) rectangle (21*\gridw,34*\gridh);
\filldraw[fill=blue!50!white] (2*\gridw,47*\gridh) rectangle (21*\gridw,48*\gridh);
\filldraw[fill=blue!50!white] (23*\gridw,5*\gridh)  rectangle (29*\gridw,6*\gridh);
\filldraw[fill=blue!50!white] (23*\gridw,19*\gridh) rectangle (29*\gridw,20*\gridh);
\filldraw[fill=blue!50!white] (23*\gridw,33*\gridh) rectangle (29*\gridw,34*\gridh);
\filldraw[fill=blue!50!white] (23*\gridw,47*\gridh) rectangle (29*\gridw,48*\gridh);
\draw [draw=black, xscale=\gridw, yscale=\gridh] (0,0)  grid (21,53);
\draw [draw=black, xscale=\gridw, yscale=\gridh] (23,0) grid (29,53);
\filldraw[fill=white, opacity=1] (0*\gridw,26*\gridh)  rectangle (21*\gridw,27*\gridh) node[align=center,right]{\scriptsize $\cdots$} ;
\filldraw[fill=white, opacity=1] (23*\gridw,26*\gridh) rectangle (29*\gridw,27*\gridh);
\draw[fill=blue!50!white]                                            (6*\gridw,-11*\gridh) rectangle (8*\gridw,-8*\gridh) (8*\gridw,-9.5*\gridh)  node[align=center,right]{\scriptsize Pilots} ;
\draw[pattern=custom north west lines,hatchspread=4pt,hatchcolor=red](12*\gridw,-11*\gridh) rectangle(14*\gridw,-8*\gridh)(14*\gridw,-9.5*\gridh) node[align=center,right]{\scriptsize SIGNAL};
\draw[fill=white,          opacity=1]                                (18*\gridw,-11*\gridh) rectangle(20*\gridw,-8*\gridh)(20*\gridw,-9.5*\gridh) node[align=center,right]{\scriptsize DATA};

\draw [->] (-0.3*\gridw,3*\gridh) -- (-0.3*\gridw,50*\gridh);
\node[align=center,rotate=90] at (-0.7*\gridw,27*\gridh){\scriptsize Subcarriers $k$} ;
\draw [->] (0*\gridw,-2*\gridh) -- (29*\gridw,-2*\gridh);
\node[align=center,rotate=0] at (14*\gridw,-4*\gridh){\scriptsize OFDM symbol $m$} ;
\draw [|<->|] (0,59*\gridh) -- (29*\gridw,59*\gridh) node[midway,above]{\scriptsize $M'$} ;
\draw [|<->|]  (2*\gridw,54*\gridh) -- (10*\gridw,54*\gridh) node[midway,above]{\scriptsize $M'_{\rm{S}}$} ;
\draw [|<->|] (11*\gridw,54*\gridh) -- (15*\gridw,54*\gridh) node[midway,above]{\scriptsize $M'_{\rm{P}}$} ;
\draw [|<->|] (16*\gridw,54*\gridh) -- (20*\gridw,54*\gridh) node[midway,above]{\scriptsize $M'_{\rm{P}}$} ;
\draw [|<->|] (24*\gridw,54*\gridh) -- (26*\gridw,54*\gridh) node[midway,above]{\scriptsize $M'_{\rm{A}}$} ;
\draw [|<->|] (27*\gridw,54*\gridh) -- (29*\gridw,54*\gridh) node[midway,above]{\scriptsize $M'_{\rm{E}}$} ;
\end{tikzpicture}
\caption{A modified 802.11p frame in subcarrier-time grid showing the position of the pilots and the data symbols.}
\label{fig:Mod80211frame}
\end{figure}
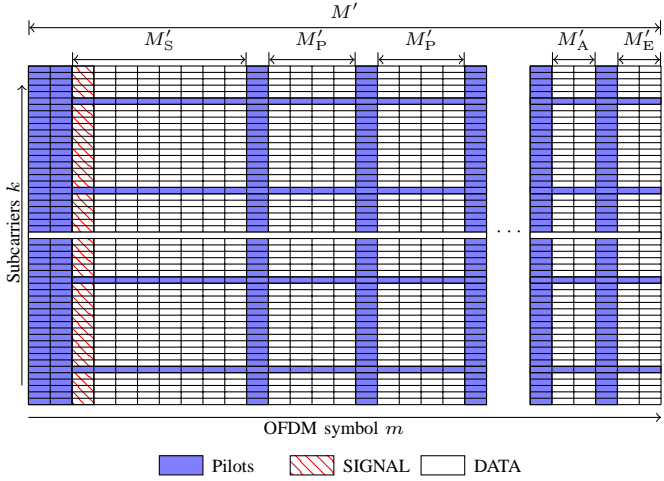

\figref{fig:Mod80211frame} shows an example of the MF with periodically inserted \GLS{OFDM} training symbols. The inserted \GLS{OFDM} symbols will henceforth be referred to as pseudo training (PT) symbols and the number of \GLS{OFDM} symbols between two periodically inserted PTs is denoted by $M'_{\rm{P}}$, which is the design parameter of the MF. $M'_{\rm{P}}$ can be fixed or made adaptive, in which case the transmitter chooses an $M'_{\rm{P}}$ and includes its value in the transmitted frame. Currently unused bits in the SERVICE field can be used to convey $M'_{\rm{P}}$ to the modified receiver. As shown in the figure, the number of OFDM symbols between the LT symbols and the first PT is denoted by $M'_{\rm{S}}$ which in some cases can be larger than $M'_{\rm{P}}$ due to the insertion of the SIGNAL symbol, SERVICE field, and \gls{MAC} header by the layers below the LLC sublayer. In fact, $M'_{\rm{S}}$ is lower bounded by $\ceil*{( N_{\rm{SERV}} + N_{\rm{MACH}} + N_{\rm{MEM}}) / N_{\rm{DBPS}}}+1$, where $\ceil*{x}$ denotes the smallest integer larger or equal to $x$. Also, depending on the length $N_{\rm{FB}}$, the frame in the end may consist of less than $M'_{\rm{P}}$ \GLS{OFDM} data symbols after the periodically inserted PTs. In that case, one additional PT is inserted after the periodically inserted training in the end of the frame. The number of OFDM symbols between the final periodically inserted PT and the additional PT is denoted by $M'_{\rm{A}}$. Since the \gls{CRC} and the TAIL bits are appended to the end of the frame by the \GLS{PHY} layer, the frame consists of $M'_{\rm{E}}$ \GLS{OFDM} symbols after the final PT as shown in the figure. $M'_{\rm{E}}$ is determined by $N_{\rm{FB}}$, $M'_{\rm{P}}$, and the \gls{MCS} used and is upper bounded by $\ceil{(N_{\mathrm{CRC}} + N_{\mathrm{MEM}})/N_{\mathrm{DBPS}}} + 1$. The number of OFDM symbols in the MF beginning with the two LT symbols is denoted by $M'$.

\begin{figure}[!t]
\renewcommand{\ScPSF}{0.6}
\psfrag{pLTb} [c][c][\ScPSF]{PTb}
\psfrag{pLTb:}[c][c][\ScPSF]{PTb:}
\psfrag{PAD}  [c][c][\ScPSF]{PAD}
\psfrag{FB}   [c][c][\ScPSF]{FB}
\psfrag{FBS}  [c][c][\ScPSF]{FBS}
\psfrag{FB1}  [c][c][\ScPSF]{$\rm{FB_1}$}
\psfrag{FB2}  [c][c][\ScPSF]{$\rm{FB_2}$}
\psfrag{FBA}  [c][c][\ScPSF]{FBA}
\psfrag{FBE}  [c][c][\ScPSF]{FBE}
\psfrag{Ter}  [c][c][\ScPSF]{Termination bits}
\psfrag{Tra}  [c][c][\ScPSF]{Training bits}
\psfrag{d}    [c][c][\ScPSF]{$\cdots$}
\psfrag{btr}  [c][c][\ScPSF]{$N_{\mathrm{DBPS}}$}
\psfrag{b6}   [c][c][\ScPSF]{$N_{\mathrm{MEM}}$}
\psfrag{NPT1}[c][c][\ScPSF]{$N_{\mathrm{S}}$}
\psfrag{NPT}[c][c][\ScPSF]{$N_{\mathrm{P}}$}
\psfrag{NPTA}[c][c][\ScPSF]{$N_{\mathrm{A}}$}
\psfrag{NPTE}[c][c][\ScPSF]{$N_{\mathrm{E}}$}
\centering
\includegraphics[width=0.95\linewidth]{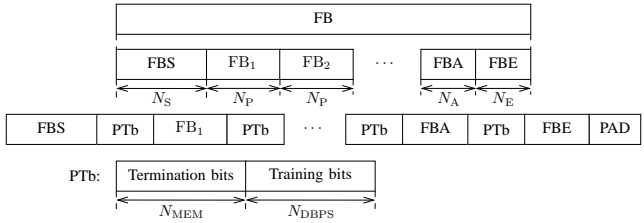}
\caption{Diagram showing the insertion of the training bits after the LLC sublayer.}
\label{fig:pLTinsertion}
\end{figure}

To insert PT symbols as shown in \figref{fig:Mod80211frame} any binary sequence of length $N_{\rm{MEM}} + N_{\mathrm{DBPS}}$ denoted by PTb is chosen which is also known at the receiver. The first $N_{\rm{MEM}}$ bits of the sequence terminate the encoder in a known state and the remaining $N_{\mathrm{DBPS}}$ bits correspond to one \GLS{OFDM} training symbol. To ensure that the inserted bits result in a complete \GLS{OFDM} symbol as shown in the figure, the insertion has to be performed at specific positions, considering the interleaver and the other fields added. Since, the scrambling is done in a random fashion unknown to the LLC layer, the choice of bits in PTb cannot be optimized. For BPSK and QPSK modulations all choices give the same performance. The procedure to insert training bits is described in procedure InsertPT and \figref{fig:pLTinsertion}. The procedure is valid for any of the \gls{MCS} specified in~\cite[Table 18.4]{IEEE802112012}. All the existing 802.11p systems can transmit the MF without undergoing any change in the \gls{MAC} and \GLS{PHY} layers.

\begin{algorithm}
\label{alg:insertpLT}
{\fontsize{8}{10}\selectfont
\begin{algorithmic}
\Procedure{InsertPT}{FB, PTb, $N_{\rm{DBPS}}$, $M'_{\rm{P}}$}
    \begin{flalign*}
    M'_{\rm{S}} & = \max (\ceil*{( N_{\rm{SERV}} + N_{\rm{MACH}} + N_{\rm{MEM}}) / N_{\rm{DBPS}}}+1 , M'_{\rm{P}}) \\
    N_{\rm{S}} &= (N_{\rm{DBPS}}\cdot (M'_{\rm{S}}-1)  )-N_{\rm{SERV}}-N_{\rm{MACH}}-N_{\rm{MEM}}
    \end{flalign*}
    \If{$N_{\rm{FB}} \geq N_{\rm{S}}$}  
        \State Group the first $N_{\rm{S}}$ bits of FB in block FBS.
        \begin{flalign*}
        \mbox{Compute: }N_{\rm{P}}  &= (N_{\rm{DBPS}}\cdot M'_{\rm{P}})-N_{\rm{MEM}} \\
        Q &= \floor*{ (N_{\rm{FB}} - N_{\rm{S}})/ N_{\rm{P}}}
        \end{flalign*}
        \If{$Q \neq 0$}
            \State Group the subsequent bits of FB into $Q$ blocks of
            \State $N_{\rm{P}}$ bits each and denote the blocks as $\rm{FB_1}, \rm{FB_2}, ..., \rm{FB_Q}$.
        \EndIf
        \\
        \If  {$ (N_{\rm{FB}} - N_{\rm{S}} - (Q \cdot N_{\rm{P}})) > N_{\rm{DBPS}}$}
            \begin{flalign*}
            M'_{\rm{A}} &= \floor*{(N_{\rm{FB}} - N_{\rm{S}} - (Q \cdot N_{\rm{P}})) /N_{\rm{DBPS}}}  \\
            N_{\rm{A}} &= (M'_{\rm{A}} \cdot N_{\rm{DBPS}}) - N_{\rm{MEM}} \\
            A &= 1
            \end{flalign*}
            \State Group the next $N_{\rm{A}}$ bits of FB into the block FBA.
        \Else
            \State $M'_{\rm{A}} = 0;  N_{\rm{A}} = 0;  A = 0$
        \EndIf
        \begin{align*}
        \mbox{Compute: }N_{\rm{E}} = N_{\rm{FB}} - N_{\rm{S}} - (Q \cdot N_{\rm{P}}) - N_{\rm{A}}
        \end{align*}
    \Else    
        \State The length of the frame is short, transmit an SF.
        \State Go to \textbf{end procedure}
    \EndIf  \\ 
\begin{itemize}
\item Group the last $N_{\rm{E}}$ bits in block FBE.
\item Insert the binary sequence PTb in between the blocks formed in the above steps as shown in \figref{fig:pLTinsertion}.
\item Pad zeros to the resulting sequence of bits if necessary to make the length of the sequence an integer multiple of octets.
\item Denote the sequence as Modified FB, which is passed down to MAC layer in~\figref{fig:dataflow}.
\end{itemize}
\EndProcedure
\end{algorithmic}
}
\end{algorithm}

\section{Receiver for Modified {802.11p} Frame}
\label{sec:blkdec}
\begin{figure*}
\renewcommand{\ScPSF}{0.8}
\psfrag{a}  [c][c][\ScPSF]{\xcancel{CP}}
\psfrag{aa} [c][c][\ScPSF]{DFT}
\psfrag{b}  [c][c][\ScPSF]{Equalizer}%
\psfrag{c}  [c][c][\ScPSF]{Demodulator}%
\psfrag{d}  [c][c][\ScPSF]{Deinterleaver}%
\psfrag{e}  [c][c][\ScPSF]{Decoder}%
\psfrag{f}  [c][c][\ScPSF]{Descrambler}%
\psfrag{g}  [c][c][\ScPSF]{Channel Estimator}%
\psfrag{rmn}[c][c][\ScPSF]{$r[m,n]$}%
\psfrag{R1} [c][c][\ScPSF]{$R[m,k]$}%
\psfrag{R2} [r][c][\ScPSF]{$R[m,k]$}%
\psfrag{S}  [c][c][\ScPSF]{$\hat{S}[m,k]$}%
\psfrag{P}  [r][c][\ScPSF]{$P[m,k]$}%
\psfrag{H}  [l][c][\ScPSF]{$\hat{H}[m,k]$}%
\centering
\includegraphics[width=0.8\linewidth]{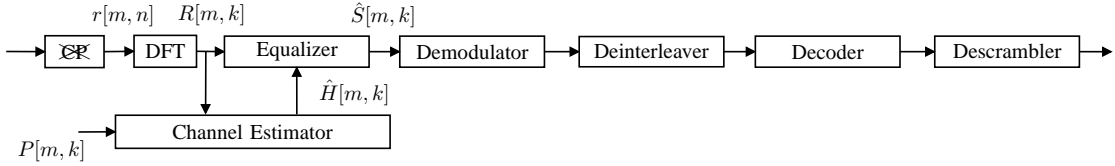}
\caption{Block diagram of a generic noniterative 802.11p receiver.}
\label{fig:gen_rec}
\end{figure*}
\begin{figure*}
\renewcommand{\ScPSF}{0.7}
\psfrag{R2}     [r][c][\ScPSF]{$R[m,k]$}%
\psfrag{P}      [r][c][\ScPSF]{$P[m,k]$}%
\psfrag{HLS}    [c][c][\ScPSF]{$\hat{H}_{\rm{P}}[m,k]$}%
\psfrag{H}      [l][c][\ScPSF]{$\hat{H}[m,k]$}%
\psfrag{LS1}    [c][c][\ScPSF]{\begin{tabular}{c@{}}LS channel\\estimation of LTs\end{tabular}}%
\psfrag{SS}     [c][c][\ScPSF]{\begin{tabular}{c@{}}Scrambler seed\\estimation\end{tabular}}%
\psfrag{Gen}    [c][c][\ScPSF]{\begin{tabular}{c@{}}Compute PT\\symbols\end{tabular}}%
\psfrag{LS2}    [c][c][\ScPSF]{\begin{tabular}{c@{}}LS channel estimation\\of all pilots\end{tabular}}%
\psfrag{LMMSE}  [c][c][\ScPSF]{\begin{tabular}{c@{}}LMMSE channel\\estimation\end{tabular}}%
\centering
\includegraphics[width=0.7\linewidth]{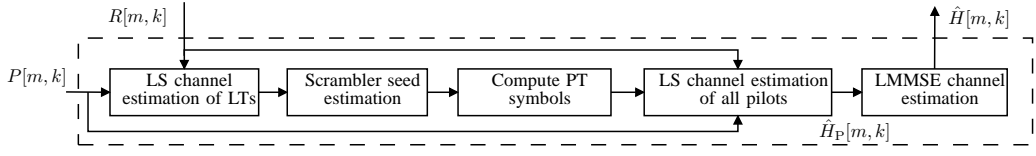}
\caption{LMMSE channel estimator for the modified frame.}
\label{fig:ChEstLMMSE}
\end{figure*}
A generic noniterative receiver for decoding the 802.11p frames is shown in \figref{fig:gen_rec}. Perfect frequency and time synchronization is assumed. The \gls{CP} of the \GLS{OFDM} symbol is discarded and an $N$-point DFT is performed to obtain the frequency domain symbols given by 
\begin{align}
R[m,k] &= H[m,k]S[m,k] + W[m,k],
\end{align}
where $H[m,k]$ is the channel frequency response at the $k$th subcarrier of the $m$th \GLS{OFDM} symbol and $W[m,k]$ are the frequency domain independent and identically distributed complex additive white Gaussian noise samples with zero mean and variance $\sigma^2$.

Very large Doppler spreads have been reported for vehicle-to-vehicle channels, which indicate very low coherence times. Coherence times in the range of $180$ to $500 \, \rm{\mu s}$ and coherence bandwidths in the range of $200$ to $700 \, \rm{kHz}$ have been reported in~\cite{Bernado2013} indicating that the vehicular channels are highly time-varying and frequency selective. This implies that we can see significant changes in the channel frequency response over the duration of an 802.11p frame. However, the channel can be considered to be approximately time-invariant over one OFDM symbol duration, and we can neglect the intercarrier interference.

The frequency domain symbols are fed to the channel estimation block. The channel estimation block uses the received symbols and the inserted pilot symbols to obtain the channel estimates $\hat{H}[m,k]$ for the whole frame. In this paper, we assume that minimum mean-squared error (MMSE) equalization is used. The output of the equalizer is therefore given by~\cite{zemen2012}
\begin{align}
\hat{S}[m,k] &=  \frac{R[m,k]\hat{H}^*[m,k]}{\sigma^2 + |\hat{H}[m,k]|^2}.
\end{align}
The equalized symbols $\hat{S}[m,k]$ are then fed to the chain of the remaining blocks (demodulator, deinterleaver, decoder, and descrambler).

The MF consists of inserted PT symbols along with the training symbols available in the SF. The available training symbols can be used in more than one way to obtain the channel estimates for the whole frame. The receiver for decoding the MF has a structure similar to the generic receiver shown in \figref{fig:gen_rec} with soft demodulator and soft input Viterbi decoder. The frequency domain symbols obtained after the DFT operation are fed to the channel estimation block that outputs the channel estimates for the whole frame. The soft demodulator outputs \glspl{LLR} of the encoded bits. The LLR values are deinterleaved and fed to the soft-input Viterbi decoder.
The channel estimation block is split into several subblocks for simplicity as shown in \figref{fig:ChEstLMMSE}. In the first step \gls{LS} channel estimates of the two LT symbols are computed and averaged to get less noisy estimates, given by
\begin{align}
\hat H_{\rm{LT}}[k] &= \frac{1}{2} \left( \frac{R[0, k]}{S[0, k]} + \frac{R[1, k]}{S[1, k]} \right).
\end{align}
The scrambler seed used to initialize the scrambler in the transmitter is necessary for reproducing the inserted PT symbols at the receiver side. The first $7$ bits in the SERVICE field correspond to the scrambler initialization. The first several DATA \GLS{OFDM} symbols are equalized with the available \gls{LS} estimates $\hat H_{\rm{LT}}[k]$ and the SERVICE field is decoded using Viterbi decoding with trace back. The number of uncoded data bits in the first DATA \GLS{OFDM} symbols chosen for decoding the SERVICE field must be larger than at least five times the constraint length requirement stated in literature to ensure reliability. Using the estimated scrambler seed, PTb binary sequence, and $M'_{\rm{P}}$, the inserted frequency domain training symbols in the PTs are generated by the receiver using the same procedure as in the transmitter. Subsequently, the \gls{LS} estimates at the LT, comb pilots, and PT positions are computed and fed to the linear minimum mean-squared error (LMMSE) estimator. The LMMSE estimator computes the channel estimates for all the positions in the frame by interpolating the \gls{LS} estimates at the pilot positions using the channel correlation matrices. The LMMSE channel estimates are then used by the MMSE channel equalizer as indicated in~\figref{fig:gen_rec}.

The \gls{LS} estimates of the channel frequency response at the pilot positions (including the PT symbols) are given by~\cite{Ovemag1996}
\begin{equation}
\hat{H}_{\rm{P}}[m,k] = \frac{R[m,k]}{P[m,k]}, \forall [m,k] \in \mathcal{P}.
\end{equation}

The computed LS estimates are arranged in a column vector ${\hat{\pmb{H}}_{\rm{P}}}$ and the channel estimates for the frame are computed using LMMSE estimation as~\cite{Ovemag1996}
\begin{equation}
\hat{\pmb{H}} = R_{ \pmb{H} \pmb{H}_{\rm{P}} }(R_{\pmb{H}_{\rm{P}}\pmb{H}_{\rm{P}}}+ \sigma^2   ( \pmb{P} \herm{\pmb{P}} )^{-1}  )^{-1} \hat{\pmb{H}}_{\rm{P}},
\end{equation}
where $\hat{\pmb{H}}$ is the vector of channel estimates at desired positions, $\pmb{P}$ is the diagonal matrix of the pilot symbols; $R_{\pmb{H}_{\rm{P}}\pmb{H}_{\rm{P}}} = \rm{E}\{\pmb{H}_{\rm{P}}  \herm{\pmb{H}_{\rm{P}}}  \}$ is the auto-correlation matrix of $\pmb{H}_{\rm{P}}$ and $R_{\pmb{H} \pmb{H}_{\rm{P}} } = \rm{E}\{\pmb{H}  \herm{\pmb{H}_{\rm{P}}} \}$ is the cross-correlation matrix between $\pmb{H}$ and $\pmb{H}_{\rm{P}}$. The elements of the correlation matrices are obtained from the channel correlation function
\begin{eqnarray}
r_H[m, m+\Delta m,k,k+ \Delta k] =\nonumber \\
\mathrm{E} \{H[m,k] H^*[m+\Delta m, k+\Delta k ]\},
\end{eqnarray}
where $\Delta m$ and $\Delta k$ are the \GLS{OFDM} symbol and subcarrier separation between the positions for which the correlation is being computed. Under the assumption that the channel is wide-sense stationary with uncorrelated scattering, the correlation function is separable and can be written as
\begin{eqnarray}
r_H[m, m+\Delta m,k,k+ \Delta k] = r_{\rm{t}}[\Delta m] r_{\rm{f}}[\Delta k], \nonumber
\end{eqnarray}
where $r_{\rm{t}}$ and $r_{\rm{f}}$ are the correlation functions in time and frequency domains, respectively. The correlation functions can be estimated using the \gls{LS} channel estimates at pilot positions or theoretically derived from the underlying channel model.

\subsection{Blockwise LMMSE Channel Estimation}
Performing the LMMSE estimation for the whole frame requires multiplication/inversion of large autocorrelation matrices, whose sizes are proportional to the length of the \GLS{OFDM} frame and number of PTs. To reduce the complexity, the LMMSE estimation can be performed block-wise. In the block-wise approach, two consecutive PTs and the symbols in between are considered as a block. LMMSE estimation for the block is performed by only considering the \gls{LS} channel estimates available in that block. This approach reduces the size of the correlation matrices to be proportional to the size of the block. The reduction in complexity is achieved at the cost of a small degradation in channel estimation in comparison to LMMSE estimation over the whole frame. Also, the LMMSE channel estimates of the second PT symbol from the blockwise LMMSE estimation are used instead of the LS estimates for the first PT of the next block, since the LMMSE estimates have a lower mean-squared error (MSE) compared to the initial LS estimates.

\subsection{Blockwise Decoding}
In an SF, the frame is terminated by $N_{\rm{MEM}}$ zero bits in a known state to reliably decode the bits at end of the frame. The positions of the inserted training bits in the MF are known in the modified receiver and can therefore be used as initial bits and terminating bits of the convolution decoder. This facilitates in decoding the frame in blocks between two PT sequences where the last $N_{\rm{MEM}}$ bits of the preceding PTb sequence serve as the initial bits and the first $N_{\rm{MEM}}$ bits of the following PTb sequence serve as the termination bits. Since the \gls{CRC} checksum is calculated for the frame including the PT sequences in the transmitter, the known PTb sequences are inserted between the block of decoded data bits to form the modified FB for which the \gls{CRC} checksum is valid. Since more bits are decoded with known termination, the overall \gls{FER} performance improves in comparison to continuous decoding with trace-back or decoding the whole frame with a single termination. It has been observed in the simulation results that the block-wise decoding offers performance gain in terms of FER.

\section{Numerical Results}
The performance of the MF with the described receiver is evaluated using computer simulations and compared with other receiver configurations decoding the SF. FER is the focus of our simulation results since it is one of the most important performance metrics for safety related applications. A frame is considered to be in error if the appended \gls{CRC} checksum does not agree with the checksum computed in the receiver. The \gls{MCS} with QPSK mapping and code rate $1/2$ in~\cite[Table 18.4]{IEEE802112012} is adopted for the safety applications and is the focus of our simulation results.

The PT symbols are inserted periodically in the MF and the choice of the period is important. Ideally the spacing between the training symbols should be proportional to the coherence time of the wireless channel. However, the safety applications require the 802.11p frames to be broadcast, meaning that a transmitted frame reaches several receivers through different channels with different channel impulse responses and relative velocities.

The channel measurement results in~\cite{Bernado2013} show that the coherence time of the vehicular channels can be as low as $180 \, \rm{\mu s}$ and this value can be much lower in case of higher relative vehicular velocities. We use $M'_{\rm{P}}=8$ \GLS{OFDM} symbols between the PT symbols, corresponding to $64 \, \rm{\mu s}$, for the simulations.

The wireless channel has been simulated with $L=15$ taps spanning the delay from $0$ to $1.4 \, \rm{\mu s}$ with a tap spacing of $0.1 \, \rm{\mu s}$. The tap gains are independent zero mean complex Gaussian with autocorrelation function $\alpha_{l} \mathrm{J_0}(2 \pi ( v / \lambda)t)$, where $v$ is the relative speed between the transmitter and the receiver, $\lambda$ is the wavelength of the electromagnetic carrier wave of frequency $f_{c}=5.9 \, \rm{GHz}$, and $\alpha_{l}$ is the power in the $l$th tap. The power in the $l$th tap is obtained from the exponentially decaying power delay profile (PDP) and is given by $\alpha_l = K \mathrm{e}^{{-\tau_l}/{\tau_{\rm{rms}}}}$, where $\tau_{l}$ is the delay of the $l$th tap, $\tau_{\rm{rms}}$ is the root-mean squared (rms) delay spread of the continuous and infinitely long exponentially decaying PDP, and $K$ is the normalization factor such that $\sum_{l=0}^{L-1} \alpha_{l} = 1$. Since, the continuous exponentially decaying PDP is truncated and sampled in our simulations, the effective rms delay spread denoted by $\tau_{\rm{erms}}$ is different from $\tau_{\rm{rms}}$. All the simulations have been performed with $\tau_{\rm{rms}} = 0.4 \, \rm{\mu s}$, which results in an effective rms delay spread of $\tau_{\rm{erms}} = 0.322 \, \rm{\mu s}$.

Theoretical channel correlation functions corresponding to the used Doppler spectra and PDP are used to generate the channel correlation matrices in the LMMSE channel estimation.

The FER results are plotted against ${E_{\rm{s}}}/{N_{\rm{0}}}$, where $E_{\rm{s}}$ is the average energy of the frequency domain symbols $S[m,k]$ and $N_{\rm{0}}/2$ is the power spectral density of the channel noise such that $N_{\rm{0}} = \sigma^2$. The rate of the frequency domain symbols $S[m,k]$ does not depend on the MCS used, the length of the frame, or if the transmitted frame is standard or modified. Hence, $E_{\rm{s}}$ is proportional to the received power with the same proportionality constant for all MCSs, frame lengths, and regardless of whether an SF or an MF is used.

\begin{figure}[!t]
\renewcommand{\ScPSF}{0.7}
\psfrag{R}   [c][c][\ScPSF]{Effective bit rate in Mbps}%
\psfrag{NFB} [c][c][\ScPSF]{Length of the FB in bytes}%
\psfrag{Std............................} [l][l][\ScPSF]{$R_{\rm{SF}}$}%
\psfrag{Mod1}[l][l][\ScPSF]{$R_{\rm{MF}}; \, M'_{\rm{P}} = 8$}%
\psfrag{Mod2}[l][l][\ScPSF]{$R_{\rm{MF}}; \, M'_{\rm{P}} = 16$}%
\centering
\includegraphics[width=\linewidth]{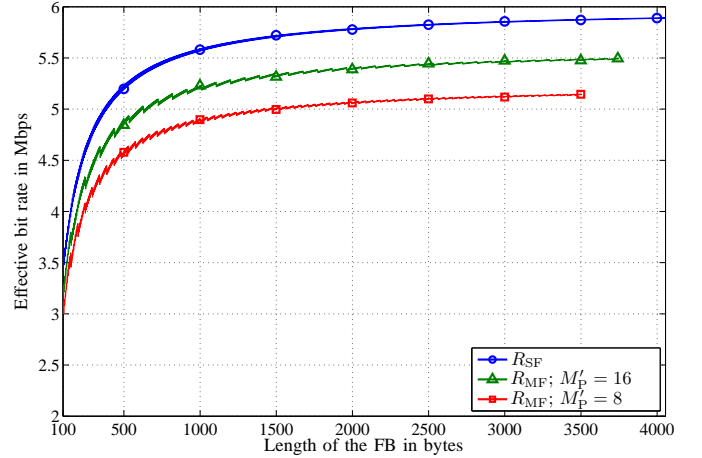}
\caption{Effective bit rates of standard and modified 802.11p frames with different spacing between the PTs.}
\label{fig:effRb}
\end{figure}

The overhead introduced by the pilot symbols and the \gls{CP} results in loss of spectral efficiency. The effective bit rate is calculated as the ratio of the number of bits in the FB and the total time duration of the frame in seconds. An MF has $(1+Q+A)$ extra OFDM symbols compared to the SF carrying same information payload. Hence, the total number of bits before the PHY layer zero padding in the SF and MF, denoted by $N_{\rm{SF}}$ and $N_{\rm{MF}}$, respectively, are
\begin{align*}
N_{\rm{SF}} &= N_{\rm{SERV}} + N_{\rm{MACH}}+ N_{\rm{FB}} + N_{\rm{CRC}} + N_{\rm{MEM}},\\
N_{\rm{MF}} &= N_{\rm{SF}} + (1+Q+A)\cdot (N_{\rm{DBPS}}+N_{\rm{MEM}}).
\end{align*}
The effective bit rates of an SF, $R_{\rm{SF}}$, and an MF, $R_{\rm{MF}}$, are computed as
\begin{align*}
R_{\rm{SF}} &= \frac{N_{\rm{FB}}}  { (5 + \ceil*{N_{\rm{SF}}/ N_{\rm{DBPS}}}) \cdot T_{\rm{SYM}}}, \\
R_{\rm{MF}} &= \frac{N_{\rm{FB}}}  { (5 + \ceil*{N_{\rm{MF}}/N_{\rm{DBPS}}}) \cdot T_{\rm{SYM}}},
\end{align*}
where the constant $5$ represents the number of OFDM symbols equivalent to the duration of the short training, LT, and SIGNAL symbols. \figref{fig:effRb} shows the effective bit rates of the SFs and the MFs for the \gls{MCS} with QPSK and code rate $1/2$. The loss in throughput when using the proposed MFs is evident from the figure. The effective bit rate curves of the MFs have been terminated when the length of the modified FB exceeds the maximum allowed length.

In the simulations the length of the FB is $N_{\rm{FB}}=(146 \cdot 8)$ bits, which results in an SF of $M=35$ \GLS{OFDM} symbols and an MF of $M'=39$ \GLS{OFDM} symbols. The MF contains one data \GLS{OFDM} symbol after the final PT at end of the frame ($A=0 ; \, M'_{\rm{E}}=1 $). The channel estimates of the final PT are used for equalizing the final data \GLS{OFDM} symbol. First $3$ DATA OFDM symbols are used to estimate the scrambler seed as described in Sec.~\ref{sec:blkdec}.

\begin{figure}[!t]
\renewcommand{\ScPSF}{0.7}
\psfrag{EsN0dB}     [c][c][\ScPSF]{${E_{\rm{s}}}/{N_{\rm{0}}}$ in dB}%
\psfrag{FER}        [c][c][\ScPSF]{FER}%
\psfrag{M1111111111111111111111}       [l][l][\ScPSF]{$\;$ SF; LTLS  }%
\psfrag{M2}       [l][c][\ScPSF]{SF; SFMMSE}%
\psfrag{M3}       [l][c][\ScPSF]{MF; MBMMSE}%
\psfrag{M4}       [l][c][\ScPSF]{MF; MFMMSE}%
\psfrag{M5}       [l][c][\ScPSF]{SF; Perfect CSI}%
\centering
\centering
\includegraphics[width=\linewidth]{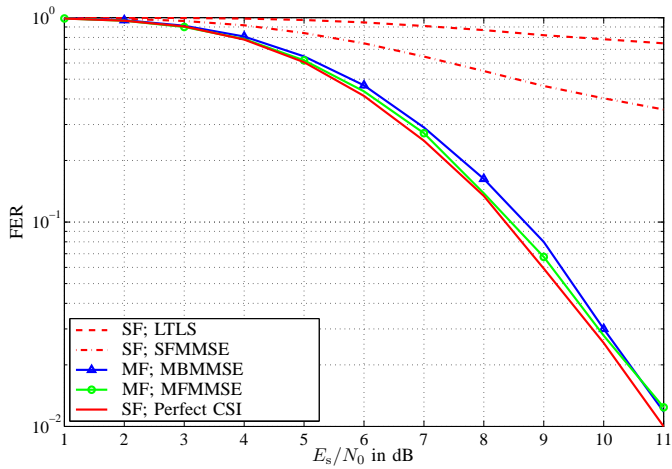}
\caption{FER of standard and modified 802.11p frames with different receiver configurations. The channel is modeled with relative vehicular velocity $v=100$  kmph. Frames carry a FB of length $N_{\rm{FB}} = (146 \cdot 8)$ bits.}
\label{fig:methods}
\end{figure}

\figref{fig:methods} shows the FER performance of the standard and modified 802.11p frames with different receiver configurations. Relative vehicular velocity $v = 100$ km/h is used for the simulations. The receiver that makes use of the \gls{LS} channel estimates of the first two LT symbols for decoding the SF (denoted by LTLS) has the worst performance. The frame LMMSE receiver for the SF (denoted by SFMMSE) makes use of all the pilots in the SF for LMMSE channel estimation and provides an improved FER performance in comparison to the LTLS receiver. The frame LMMSE with block decoding receiver for the MF (denoted by MFMMSE) is close to the performance of SF with perfect channel state information (CSI). The low complexity block LMMSE with block decoding receiver for the MF (denoted by MBMMSE) has slightly worse FER compared to the MFMMSE receiver. To summarize, inserting PT symbols and using them efficiently can improve the FER performance significantly and performance close to perfect channel estimation can be obtained.

\begin{figure}[!t]
\renewcommand{\ScPSF}{0.65}
\psfrag{EsN0dB}     [c][c][\ScPSF]{${E_{\rm{s}}}/{N_{\rm{0}}}$ in dB}%
\psfrag{FER}        [c][c][\ScPSF]{FER}%
\psfrag{M111111111111111111111111111111111111111111111111}[l][l][\ScPSF]{$\;$ $N_{\rm{FB}}=(335 \cdot 8)$; $v = 100$ kmph; MF; MBMMSE}%
\psfrag{M2}       [l][c][\ScPSF]{$N_{\rm{FB}}=(335 \cdot 8)$; $v = 100$ kmph; SF ; Perfect CSI}%
\psfrag{M3}       [l][c][\ScPSF]{$N_{\rm{FB}}=(146 \cdot 8)$; $v = 200$ kmph; MF; MBMMSE}%
\psfrag{M4}       [l][c][\ScPSF]{$N_{\rm{FB}}=(146 \cdot 8)$; $v = 200$ kmph; SF ; Perfect CSI}%
\psfrag{M5}       [l][c][\ScPSF]{$N_{\rm{FB}}=(146 \cdot 8)$; $v = 100$ kmph; MF; MBMMSE}%
\psfrag{M6}       [l][c][\ScPSF]{$N_{\rm{FB}}=(146 \cdot 8)$; $v = 100$ kmph; SF ; Perfect CSI}%
\centering
\includegraphics[width=\linewidth]{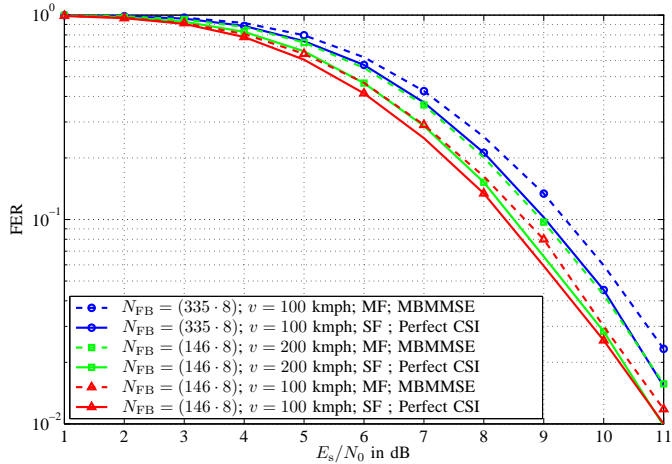}
\caption{FER of standard and modified 802.11p frames for different packet sizes and relative vehicular velocities using block LMMSE channel estimation with block decoding.}
\label{fig:velocity}
\end{figure}

In \figref{fig:velocity} the FER performance of the MF with different $N_{\rm{FB}}$ and relative vehicular velocities is benchmarked against the SF with perfect CSI. MBMMSE receiver configuration is used for obtaining the MF results. From the figure it can be observed that the FER performance of the MF does not degrade for long packets and high vehicular velocities due to the periodic nature of the inserted PT symbols, and is close to the performance of SF with perfect CSI. Moreover, the loss against the idealized case (SF; Perfect CSI) is less than $0.5$ dB in all the scenarios.

\begin{figure}[!t]
\renewcommand{\ScPSF}{0.7}
\psfrag{EsN0dB}     [c][c][\ScPSF]{${E_{\rm{s}}}/{N_{\rm{0}}}$ in dB}%
\psfrag{FER}        [c][c][\ScPSF]{FER}%
\psfrag{M11111111111111111111111111111111111111111}[l][l][\ScPSF]{$\,$ MF; $N_{\rm{FB}}=(146 \cdot 8)$; SFMMSE; $\;\; M'_{\rm{P}}=8$}%
\psfrag{M2}       [l][c][\ScPSF]{SF; $\; N_{\rm{FB}}=(146 \cdot 8)$; SFMMSE}%
\psfrag{M3}       [l][c][\ScPSF]{MF; $N_{\rm{FB}}=(147 \cdot 8)$; MBMMSE; $M'_{\rm{P}}=16$}%
\psfrag{M4}       [l][c][\ScPSF]{MF; $N_{\rm{FB}}=(146 \cdot 8)$; MBMMSE; $M'_{\rm{P}}=8$}%
\psfrag{M5}       [l][c][\ScPSF]{SF; $\; N_{\rm{FB}}=(146 \cdot 8)$; Perfect CSI}%
\centering
\includegraphics[width=\linewidth]{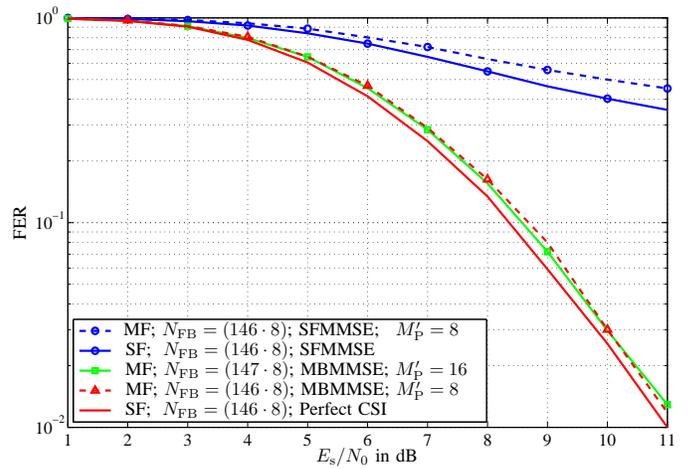}
\caption{FER of standard and modified 802.11p frames. The channel is modeled with relative vehicular velocity $v=100$  kmph. The effect of decoding the MF with standard receiver is seen in the FER curves. PT separation of $M'_{\rm{P}} = 16$ does not affect the FER for the chosen channel.}
\label{fig:final}
\end{figure}

\figref{fig:final} shows the FER results when
(i) an MF is decoded with SFMMSE, i.e., when the PTs are not used for channel estimation (circles). The loss in performance is due to the fact that MF is longer than the SF for the same length of FB; (ii) a larger $M'_{\rm{P}} \, (M'_{\rm{P}} = 16)$ is chosen. A larger separation of PTs with $M'_{\rm{P}} = 16$ does not affect the channel estimation for the chosen channel as seen from the figure (squares) and the FER performance coincides with the performance when $M'_{\rm{P}} = 8$.

\section{Conclusion}
In this paper a method to insert additional training symbols in an 802.11 OFDM frame that works for all MCSs and frame lengths has been proposed. We have also shown a method of informing the receiver about the period of the inserted training symbols through the use of unused bits in the SERVICE field. Two algorithms for decoding the MF have also been described to exploit the
additional training symbols inserted.

The main pros and cons of our method are: (i) The transmitter PHY and MAC layers do not need modification to transmit the MF (with the exception of utilizing the currently unused bits in SERVICE field); (ii) The described receiver for MFs can perform close to the case with perfect CSI, without decision feedback and channel tracking, even for channels with fast time-variations; (iii) A legacy receiver will also be able to decode an MF at the price of a small performance loss; (iv) The additional training symbols result in loss of spectral efficiency.

Safety related applications are being widely tested using software defined radio implementations of 802.11p. The proposed MF and receiver can be easily implemented starting from the 802.11p implementations. The results show that periodically inserted \GLS{OFDM} training symbols in an 802.11p frame can provide good FER performance with a noniterative receiver and as a consequence such a pattern of training data can be a possibility for the future revisions of 802.11p for vehicular communications.

\bibliography{Bibliography}

\begin{thebibliography}{1}
\providecommand{\url}[1]{#1}
\csname url@samestyle\endcsname
\providecommand{\newblock}{\relax}
\providecommand{\bibinfo}[2]{#2}
\providecommand{\BIBentrySTDinterwordspacing}{\spaceskip=0pt\relax}
\providecommand{\BIBentryALTinterwordstretchfactor}{4}
\providecommand{\BIBentryALTinterwordspacing}{\spaceskip=\fontdimen2\font plus
\BIBentryALTinterwordstretchfactor\fontdimen3\font minus
  \fontdimen4\font\relax}
\providecommand{\BIBforeignlanguage}[2]{{%
\expandafter\ifx\csname l@#1\endcsname\relax
\typeout{** WARNING: IEEEtran.bst: No hyphenation pattern has been}%
\typeout{** loaded for the language `#1'. Using the pattern for}%
\typeout{** the default language instead.}%
\else
\language=\csname l@#1\endcsname
\fi
#2}}
\providecommand{\BIBdecl}{\relax}
\BIBdecl

\bibitem{Bernado2013}
L.~Bernad\'o, T.~Zemen, F.~Tufvesson, {A. F. Molisch}, and C.~Mecklenbr\"auker,
  ``Delay and {Doppler} spreads of non-stationary vehicular channels for safety
  relevant scenarios,'' \emph{IEEE Transactions on Vehicular Technology},
  no.~99, 2013.

\bibitem{zemen2012}
T.~Zemen, L.~Bernad\'o, N.~Czink, and {A. F. Molisch}, ``Iterative time-variant
  channel estimation for 802.11p using generalized discrete prolate spheroidal
  sequences,'' \emph{IEEE Transactions on Vehicular Technology}, vol.~61,
  no.~3, pp. 1222--1233, 2012.

\bibitem{Woong2009}
W.~Cho, S.~I. Kim, H.-K. Choi, H.-S. Oh, and D.-Y. Kwak, ``Performance
  evaluation of {V2V/V2I} communications: The effect of midamble insertion,''
  in \emph{1st International Conference on Wireless Communication, Vehicular
  Technology, Information Theory and Aerospace Electronic Systems Technology},
  2009, pp. 793--797.

\bibitem{IEEE802112012}
``Wireless {LAN} medium access control ({MAC}) and physical layer ({PHY})
  specifications,'' \emph{IEEE Std 802.11-2012}, pp. 1--2793, 2012.

\bibitem{IEEE80211p}
``Wireless {LAN} medium access control ({MAC}) and physical layer ({PHY})
  specifications, {Amendment 6}: Wireless access in vehicular environments,''
  \emph{IEEE Std 802.11p-2010}, pp. 1--51, 2010.

\bibitem{Ovemag1996}
O.~Edfors, M.~Sandell, J.-J. van~de Beek, {S. K. Wilson}, and {P. O.
  B\"orjesson}, ``{OFDM} channel estimation by singular value decomposition,''
  in \emph{IEEE 46th Vehicular Technology Conference}, vol.~2, 1996, pp.
  923--927.

\end{thebibliography}

\end{document}